\documentclass[a4paper,epj,nopacs]{svjour}
\usepackage[reqno]{amsmath}
\usepackage{amsfonts,amssymb,latexsym}
\usepackage{graphicx}

\newcommand{\m}{\hphantom{$-$}}

\newcommand{\psibar}{\bar{\psi}}
\newcommand{\half}{\frac{1}{2}}
\newcommand{\bra}{\langle}
\newcommand{\ket}{\rangle}
\newcommand{\pbp}{\bra\psibar\psi\ket}
\newcommand{\var}{\operatorname{var}}

\newcommand{\U}{\operatorname{U}}

\newcommand{\ctsmb}{\cite{Montvay:1996ea,Montvay:1999ty}}
\newcommand{\cadj}{\cite{Hands:2000ei}}
\newcommand{\ckstvz}{\cite{Kogut:2000ek}}
\newcommand{\ccsc}{\cite{Bailin:1984bm,Alford:1997zt,Rapp:1998zu,Alford:1998mk}}
\newlength{\colw}
\begin{document}

\title{Diquark condensation in dense adjoint matter}

\author{Simon Hands\inst{1} \and Istv\'an Montvay\inst{2} \and Luigi
Scorzato\inst{1} \and Jonivar Skullerud\inst{2}}

\institute{Department of Physics, University of Wales Swansea,
 Singleton Park, Swansea SA2 8PP, UK
\and Theory Division, DESY, Notkestra{\ss}e 85, D-22603 Hamburg, 
Germany}

\abstract{
We study SU(2) lattice gauge theory at non-zero chemical potential
with one staggered quark flavor in the adjoint representation.  In
this model the fermion determinant, although real, can be both
positive and negative.  We have performed numerical simulations using
both  hybrid Monte Carlo and two-step multibosonic algorithms, the
latter being capable of exploring sectors with either determinant
sign.  We find that the positive determinant sector behaves like a
two-flavor theory, with the chiral condensate rotating into a
two-flavor diquark condensate for $\mu>m_\pi/2$, implying a superfluid
ground state. Good agreement is found with analytical predictions made
using chiral perturbation theory. In the `full' model there is no sign
of either onset of baryon density or diquark condensation for the
range of chemical potentials we have considered.  The impact of the
sign problem has prevented us from exploring the true onset transition
and the mode of diquark condensation, if any, for this model.
} \maketitle

\section{Introduction}

In recent years, significant progress has been made in understanding
the phase diagram of QCD at non-zero baryon chemical potential (for
reviews, see \cite{Rajagopal:2000wf,Alford:2001dt}).  On the basis of
numerous model calculations, it is now believed that the ground state
of QCD at high density is characterised by a diquark condensate which
spontaneously breaks gauge and/or baryon number symmetries \ccsc.
However, although the results appear to be qualitatively independent
of the specific model and approximation employed, little can be said
quantitatively due to the lack of a first-principles, nonperturbative
method that can access the relevant regions of the phase diagram.

Lattice QCD, which would be such a method, fails because the
Euclidean-space fermion determinant becomes complex when a chemical
potential is included, so standard algorithms cannot be applied.
However, it is possible to study QCD-like theories where the fermion
determinant remains real even at non-zero $\mu$.  These theories can
be used as testbeds to examine the validity of the models used to
study real QCD, as well as directly to improve our understanding of
phenomena such as diquark condensation and phase transitions in dense
matter.  Examples of such theories are two-color QCD, QCD with adjoint
quarks or at non-zero isospin chemical potential \cite{Son:2000xc},
and the Nambu--Jona-Lasinio model
\cite{Hands:1998kk,Lucini:2000}.  A further point of
interest is that chiral perturbation theory ($\chi$PT) may also be
 applied to many of these theories \cite{Kogut:1999iv,Kogut:2000ek,Splittorff:2000mm}.

Two-color QCD with a baryon chemical potential
has been an object of study for lattice theorists for many years
\cite{Nakamura:1984,Dagotto:1987,klatke:1990};
there have recently been a number of simulations 
with quarks in the fundamental representation
\cite{Morrison:1998ud,Hands:1999md,Aloisio:2000if,Bittner:2000rf,Liu:2000in,Muroya:2000qp,Kogut:2001na,Kogut:2001if}.
Here we will be studying two-color QCD with staggered fermions in the
adjoint representation.  The symmetries and conjectured phase diagram
of this model have been presented in a previous paper \cadj; here we
summarise the main features:
\begin{itemize}
\item For an odd number of staggered flavors $N$, the fermion determinant may be
negative.  This means that this model still has a sign problem, which
may make simulations at large $\mu$ difficult.
\item At $m=\mu=0$ the $\U(N)\times\U(N)$ flavor symmetry is enhanced
to a $\U(2N)$ symmetry which relates quarks to antiquarks.  This
symmetry is broken by the chiral condensate to $\mathrm{Sp}(2N)$,
leaving $N(2N-1)$ massless Goldstone modes, which become degenerate
pseudo-Goldstone states for $m\neq0$.
\item When $N\geq2$ these pseudo-Goldstone states include gauge
invariant scalar diquarks, which are thus
degenerate with the pion at $\mu=0$.  These models can be
studied for $\mu\not=0$ by $\chi$PT \ckstvz, the main result being
that for $\mu>m_\pi/2$ the chiral condensate rotates into a diquark
condensate (the two being related by a $\U(2N)$ rotation) while the
baryon density increases from zero.  For $N=2$ the diquark operator 
which condenses is
\begin{equation}
qq_{\bf3}=\frac{i}{2}\left[\chi^{p\,tr}(x)\varepsilon^{pq}\chi^q(x)+
\bar\chi^p(x)\varepsilon^{pq}\bar\chi^{q\,tr}(x)\right],
\label{eq:qq3}
\end{equation}
where $p,q=1,2$ are explicit flavor indices and $\varepsilon$ the antisymmetric
tensor.

\item The model with $N=1$ is not expected to contain any diquark
pseudo-Goldstones and is not accessible to $\chi$PT.  We expect an
onset transition as some $\mu_o\approx m_b/n_q>m_\pi/2$,
where $m_b$ is the mass of the lightest baryon and $n_q$ its baryon
charge. 
\item For $N=1$ the operator (\ref{eq:qq3}) is forbidden by the Exclusion
Principle; there is, however, a possibility of a gauge non-singlet, and hence
color superconducting,
diquark condensate at large chemical potential:
\begin{equation}
qq_{sc}^i=
\half\left[\chi^{tr}(x)t^i\chi(x)+\bar\chi(x)t^i\bar\chi^{tr}(x)\right]
\, ,
\label{eq:qqsc}
\end{equation}
where $t^i$ is a group generator in the adjoint representation.
Since $qq_{sc}^i$ also transforms in the adjoint representation of the gauge
group, the difference between the sum of the Casimirs of the constituents and
that of the composite is positive, and hence the interaction 
due to one-gluon exchange is attractive in this channel.
\end{itemize}

The indication in \cadj\ (and the update in \cite{Hands:2000yh}) is
that the positive determinant sector of the $N=1$ model behaves like
the model with $N=2$. Ignoring the determinant sign has the effect of
introducing extra
`conjugate quark' degrees of freedom $q^c$, which carry positive baryon charge
but transform in the conjugate representation of the gauge group
\cite{Stephanov:1996ki}. In three-color
QCD this approximation leads to unphysical
light $qq^c$ states which distort the physics of $\mu\not=0$ beyond 
recognition. In two-color QCD the effect is more subtle; generically the 
physical spectrum contains $qq^c$ states unless expressly forbidden by the 
Exclusion Principle. For models with staggered lattice fermions this is the case
for $N=1$ adjoint flavor \cadj. In principle, therefore, by simulating this
model we may get information about two `physical' models for the price
of one; $N=1$ by taking account of the determinant sign, and $N=2$ by ignoring
it. The present paper aims to strengthen the evidence for this scenario.
As discussed in \cadj, for $N=1$ staggered adjoint quarks, corresponding to four
physical flavors, the continuum limit is problematic since the
model is not asymptotically free. Our primary interest remains
in studying a strongly
interacting model with the potential to show superconducting behaviour
regardless of these issues.

The remainder of the article is organised as follows.  In
section~\ref{sec:algo} we study the performance of 
the two-step multibosonic algorithm.  In
section~\ref{sec:results} we present our results.  In
sec.~\ref{sec:chiral} we study the chiral condensate, fermion density,
and pion mass and susceptibilities, and demonstrate the different
physical behaviour of the positive determinant sector and the full
theory.  In sec.~\ref{sec:gauge} we study pure gauge observables and
the effect of Pauli blocking in the positive determinant sector.  
Secs.~\ref{sec:algo},~\ref{sec:chiral} and~\ref{sec:gauge}
all build on the results of \cadj.
In sec.~\ref{sec:diquark} we present new results for diquark condensation, 
in both superfluid (i.e.\ gauge singlet) and superconducting (i.e.\ gauge
non-singlet) channels.
Our conclusions are presented in section~\ref{sec:conclusions}. 

\section{Algorithm results}
\label{sec:algo}

The two-step multibosonic (TSMB) algorithm \ctsmb\ contains 6 tunable
parameters: the polynomial orders $n_1$ and $n_2$\footnote{The
orders $n_3, n_4$ must merely be chosen large enough that they give
rise only to negligible errors.}; and the number of scalar heatbath,
scalar overrelaxation, gauge metropolis and noisy correction steps
$I_H, I_O$, $I_M, I_c$ in each update cycle.  We have not systematically
explored the effect of varying the relative values of $I_H$ and $I_O$,
but we do not believe that this would have a major impact.  In
practice, we have also set $I_c=1$ which leaves us with 4 parameters
to tune.

\begin{figure*}
\begin{center}
\includegraphics[width=0.3\textwidth]{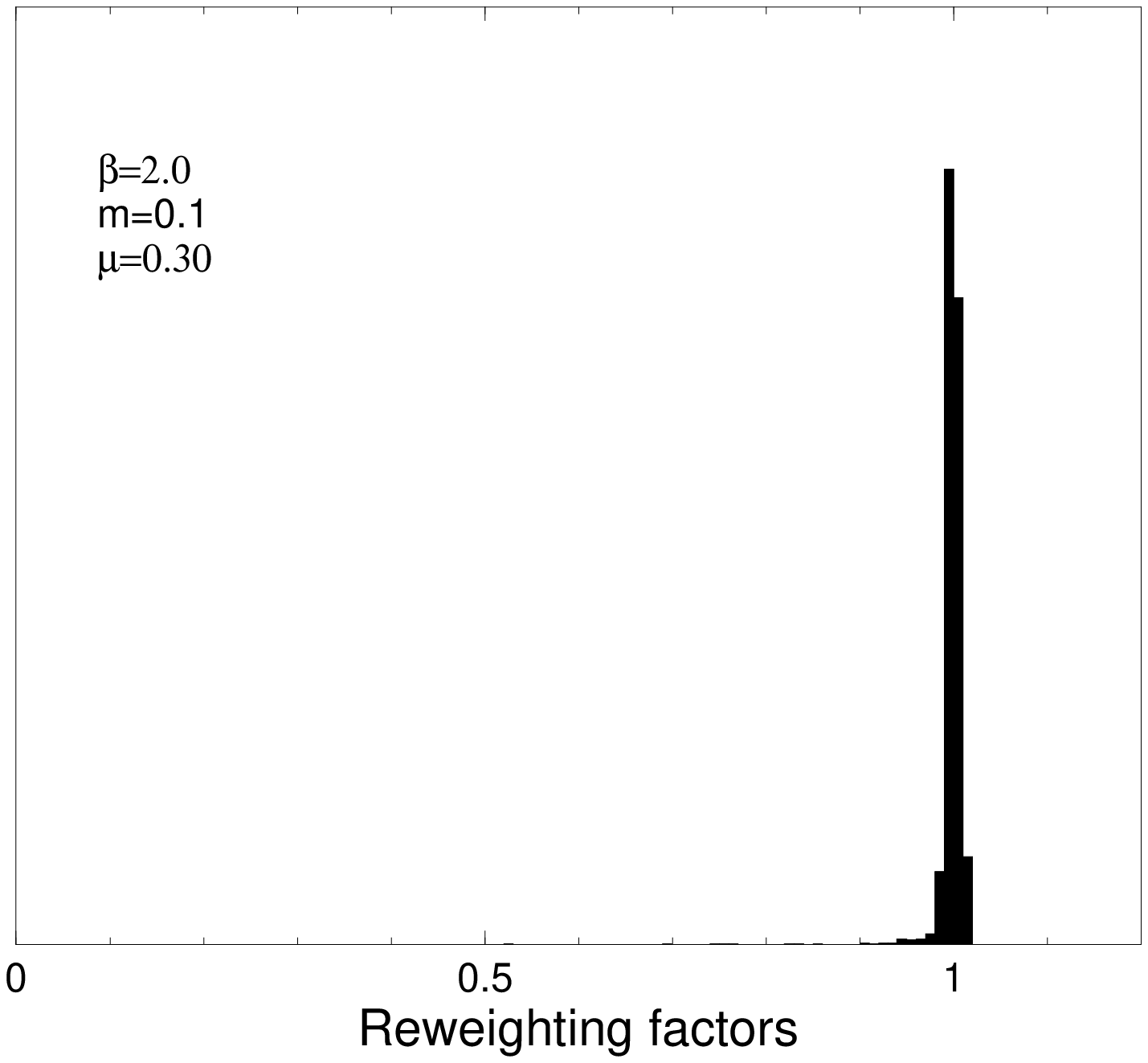}
\includegraphics[width=0.3\textwidth]{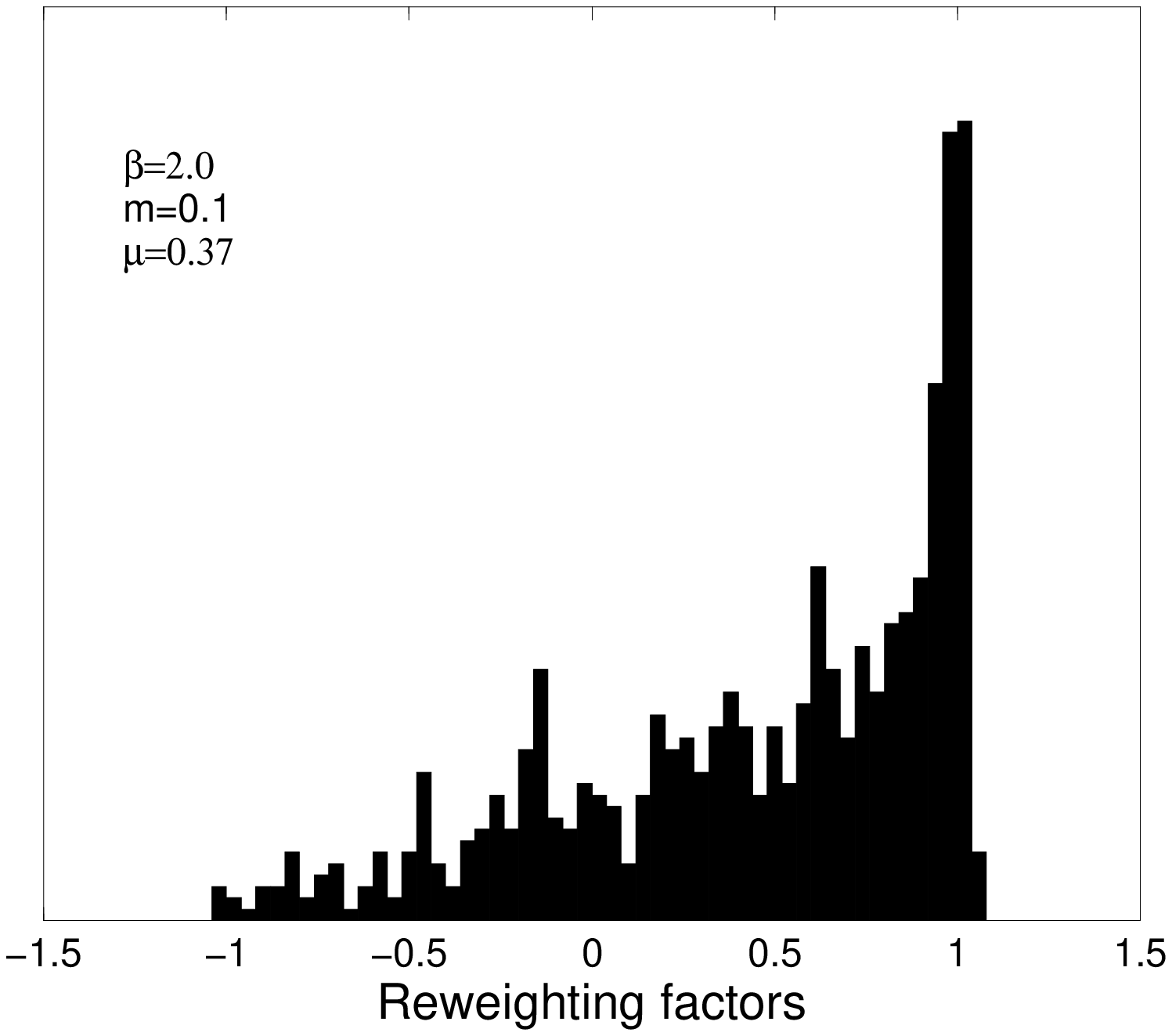}
\includegraphics[width=0.3\textwidth]{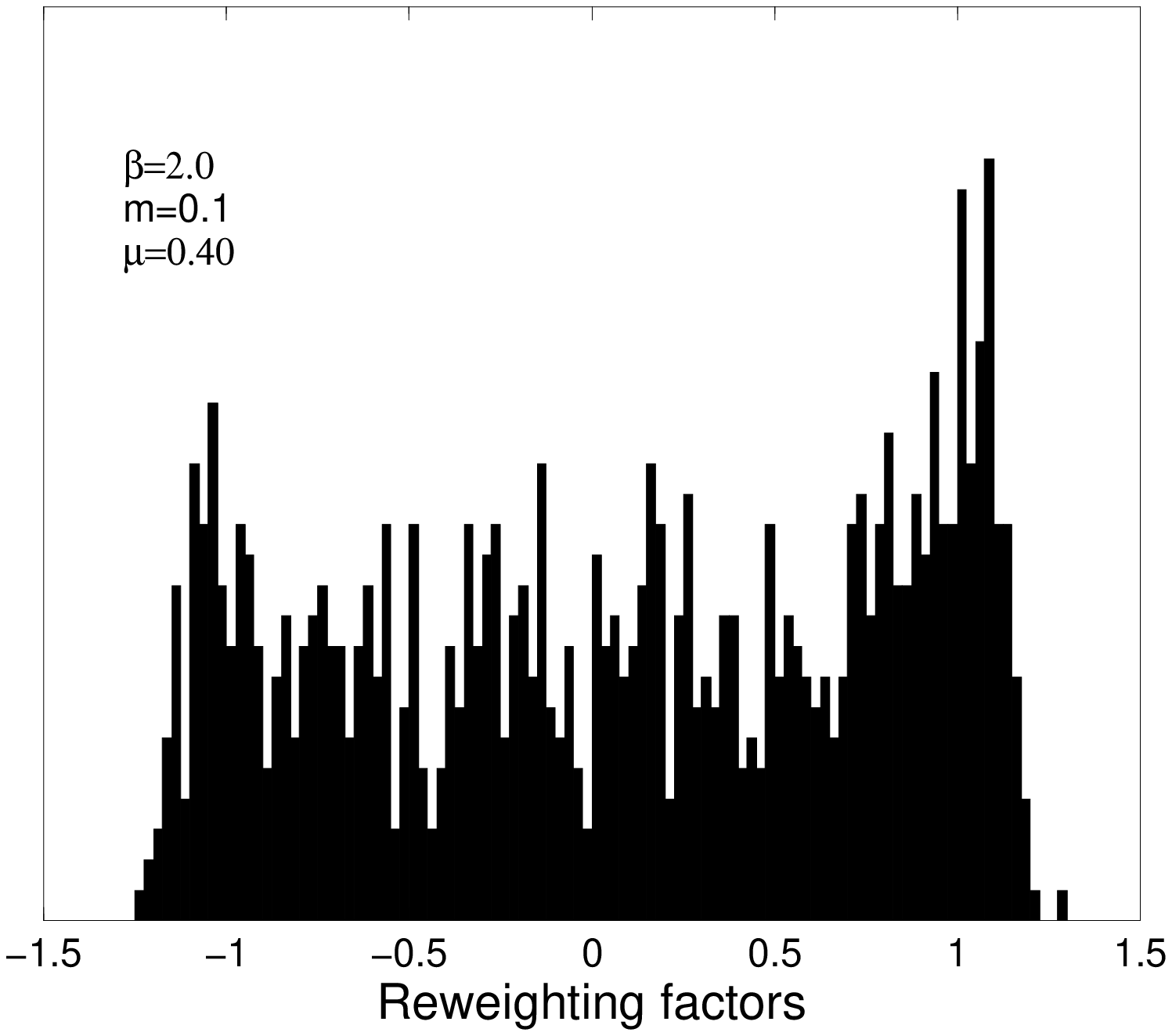}
\end{center}
\caption{Reweighting factors, for $\mu=0.3$ (left), $\mu=0.37$
(middle) and $\mu=0.4$ (right).  As the proportion of negative
determinant configurations increases, the preferred distribution of
reweighting factors changes from being sharply peaked around 1 to
nearly flat.  The polynomial orders $(n_1,n_2)$ are (48,500), (80,800)
and (100,1000) respectively.}
\label{fig:reweight}
\end{figure*}

The number of multiplications by $M^\dagger M$, where $M$ is the fermion matrix,
in a TSMB cycle is given by
\begin{equation}
n_{\text{TSMB}} \simeq \frac{7}{2}n_1(I_H+I_O+I_M) + (n_2+n_3)I_c\, .
\label{eq:matrixmult}
\end{equation}
The parameters $n_1$ and $n_2$ are essentially given by the condition
number for a typical configuration at the simulation point.  $n_1$ is
tuned to give a reasonable acceptance rate, while $n_2$ is tuned to
give reasonable reweighting factors.  What is `reasonable' in the
latter case depends on whether we are in the phase where the
determinant is positive, or whether there is a significant proportion
of negative determinant configurations.  In the former case, it is
desirable to choose $n_2$ so large that the reweighting factors are
very close to 1.  In the latter case, the sign of the determinant will
multiply the reweighting factor in the reweighting step, so a sharp
peak around 1 will also give a sharp peak around $-1$.  Instead, it is
preferable to have a flatter distribution, thus permitting the
algorithm to change the determinant sign.  This is illustrated in
fig.~\ref{fig:reweight}, where we show the reweighting factors
obtained on a $4^3\times8$ lattice with $\beta=2.0,m=0.1$ at three
values for the chemical potential: $\mu=0.3$, where the determinant is
always positive; $\mu=0.37$, where we find that 22\% of the
configurations have a negative determinant; and $\mu=0.4$, where
nearly half the configurations have a negative determinant.  In all
cases, we find that $n_2\sim10n_1$.  The optimal value of $n_1$
increases somewhat with $\mu$ (and the condition number).

\begin{figure}
\begin{center}
\includegraphics[angle=-90,width=\colw]{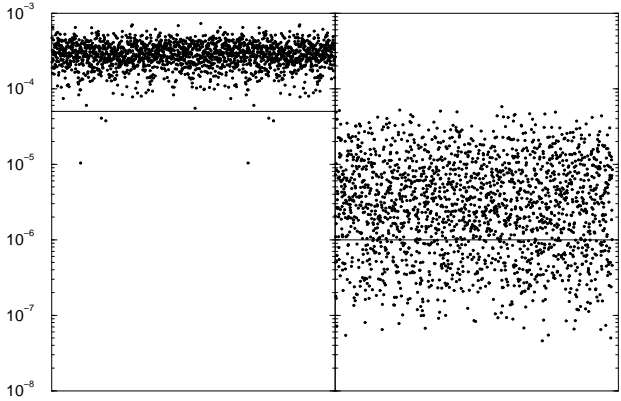}
\end{center}
\caption{The lowest eigenvalues of the hermitean fermion matrix, for
$\mu=0.3$ (left) and $\mu=0.4$ (right).  The lines indicate the lower
bound $\epsilon$ of the polynomial approximation employed.}
\label{fig:eigenvals}
\end{figure}

In fig.~\ref{fig:eigenvals} we show the distribution of the lowest
eigenvalues of the hermitean fermion matrix used in our update
procedure at $\mu=0.3$ and 0.4.  We see that the typical lowest
eigenvalue, and thereby also the condition number, changes by more
than two orders of magnitude between these two points.  Also included
is the lower limit $\epsilon$ of the polynomial approximation we have
used at these two points.  We find that at $\mu=0.3$ the polynomial
approximation is nearly always accurate, while at $\mu=0.4$ this is no
longer the case for a substantial proportion of our configurations.

How the autocorrelation depends on the values of $I_H$ and $I_M$ is
not obvious.  It turns out that the acceptance rates are virtually
unchanged when $I_M$ is changed.  This means that it should be
preferable to perform a long series of metropolis updates before the
noisy correction, since the configuration will on average change
more during the update sequence.  This becomes inefficient at the
point where the time taken for the metropolis updates begins to
dominate over the time required for the noisy correction.

To determine the autocorrelation and its dependence on the algorithm
parameters more precisely, we have performed long runs on a
$4^3\times8$ lattice with $\beta=2.0,m=0.1$ at $\mu=0.3$ and $\mu=0.37$.  The
parameters are given in table~\ref{tab:autoparams}.  A measure of how
expensive one update cycle (sweep) is, is given by the number
$N_{4hr}$ of sweeps in one 4-hour job on the Cray T3E, where this
simulation was performed.  We can see that the ratios of these numbers
do not quite match those derived from (\ref{eq:matrixmult}); this may
be attributed to the time the TSMB algorithm spends on operations
that have not been included in the
estimate (\ref{eq:matrixmult}).

\begin{table*}[htb]
\begin{center}
\begin{tabular}{rrrrrrrrrrr}
\hline
$\mu$ & $n_1$ & $n_2$ & $I_M$ & $n_{\text{sweep}}$
& $n_{\text{TSMB}}$ & $N_{4hr}$ & $\tau$ & $\tau_M$ \\ \hline
0.30 & 48 & 500 & 4  & 80080 &  3452 & 370 & 2000--3000 & \\
0.30 & 48 & 500 & 20 & 91080 &  6140 & 230 & $1500$ & $9.2\times10^6$ \\
0.37 & 80 & 800 & 4  & 96910 &  5600 & 220 & 2000--4500 & \\
0.37 & 80 & 800 & 20 & 73080 & 10080 & 140 & $\sim2500$ &
$25\times10^6$ \\
\hline
\end{tabular}
\end{center}
\caption{Parameters for our autocorrelation runs.  In all cases we
have used $I_H=2; I_O=8; I_c=1$.  $n_{\text{sweep}}$ is the length of the runs in
update cycles; $N_{4hr}$ is the number of such update cycles achieved
in 4 hours on the Cray T3E.  $n_{\text{TSMB}}$ is the number of matrix
multiplications in one update cycle according to
(\protect\ref{eq:matrixmult}).  $\tau$ is the autocorrelation time in
update cycles, while $\tau_M$ is the autocorrelation time in matrix
multiplications.}
\label{tab:autoparams}
\end{table*}

\begin{figure*}
\begin{center}
\includegraphics[angle=-90,width=0.67\textwidth]{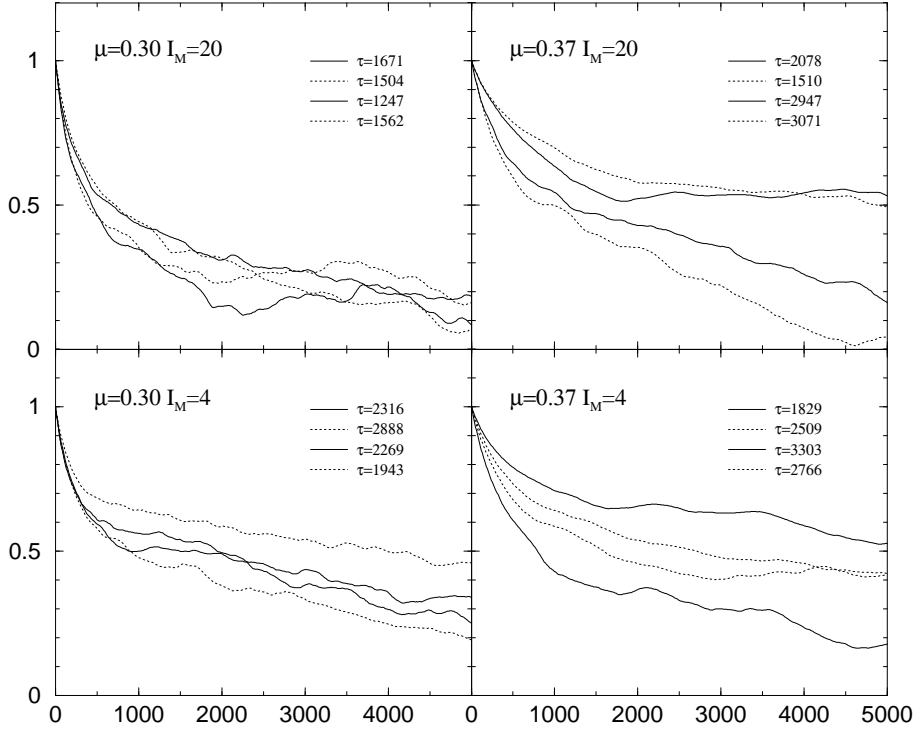}
\end{center}
\caption{Autocorrelations for our four runs.  The solid lines are the
autocorrelations of the spatial plaquette; the dotted lines are the
timelike plaquette autocorrelations.  All these runs were performed
independently on two processors, so there are two autocorrelation
curves for each quantity.}
\label{fig:autocorr}
\end{figure*}

In all cases we have updated two separate lattices, and the
autocorrelations have been measured separately for the timelike and
spacelike plaquette for each lattice.  The autocorrelation time has
been measured by two methods: firstly, by a straightforward
measurement of the autocorrelation function, giving the integrated
autocorrelation time $\tau_a$; and secondly, by determining the
jackknife variance of the plaquette with successively larger bin
sizes, until a plateau is reached.  The autocorrelation time is then
obtained as $\tau_j\approx \var(n_{\max})/2\var(1)$ where $\var(n)$ is
the jackknife variance with bin size $n$.  In
figure~\ref{fig:autocorr} we show the autocorrelation functions for
each of our four runs.  We can see that for $I_M=4$ there is a very
long tail, and the autocorrelation itself is poorly determined.  This
reflects the presence of some very slow modes, and our runs are not
long enough to accurately determine the autocorrelation in this case.
The situation is substantially improved for $I_M=20$, although there
are still substantial uncertainties and possibly residual slow modes
in the data for $\mu=0.37$, where we have only been able to obtain
order-of-magnitude estimates.  We have combined all 8 estimates in one
number or range for each parameter value; those are given in
table~\ref{tab:autoparams}.

We have also studied the autocorrelation of fermion quantities such as
the chiral condensate and the pion propagator at $\mu=0.3$.  Here we
found an autocorrelation time of $\tau_{\pbp}\sim\tau_{\pi}\sim120$
with $I_M=4$, at least an order of magnitude smaller than for the
plaquette.  This means that we can be confident that the long
plaquette autocorrelation times will not give rise to any additional
uncertainty in fermionic observables.

\section{Physics results}
\label{sec:results}

\subsection{Simulation parameters}

As in \cadj, we have used three different masses using HMC at $\beta=2.0$
on a $4^3\times8$ system, 
exploring values of $\mu$ up to and including 0.8 for $m=0.1$, $\mu=0.7$ for
$m=0.05$, and $\mu=0.5$ for $m=0.01$ \cite{Hands:2000yh}. In addition we have
performed further high statistics runs at $m=0.1$ in the region of the onset
phase transition at $\mu_o\simeq0.35$.
This latter parameter space 
was also explored using the TSMB
algorithm, permitting an assessment of the impact of the determinant sign
fluctuations.
The simulation parameters in this region for both algorithms are
given in table~\ref{tab:sim-params}. Note that in order to maintain a reasonable
acceptance rate extremely short HMC trajectory lengths are required for
$\mu>\mu_o$. We also performed an analysis of 
diquark measurables to be described in sec.~\ref{sec:diquark}.

\begin{table*}
\begin{center}
\begin{tabular}{rrrrrrrrrrr}
\hline
$\mu$ & $n_1$ & $n_2$ & $n_4$ & $N_{\text{cfg}}$
& $N_{\text{gauge}}$ & $p_-$ & $\bra r\ket$ 
& $N_{\text{traj}}$ & $\tau$ \\ \hline
0.00 &  16 &  120 &  240 & 1248 & 380 & 0.00 & 1.00 &  &  \\
0.20 &     &      &      &      &     &      &           & 2000  & 0.5 \\
0.30 &  48 &  500 &  800 & 2592 & 216 & 0.00 & 0.9982(3) & 2000  & 0.465\\ 
0.35 &     &      &      &      &     &      &           & 4000  & 0.18 \\
0.36 &  64 &  700 &  900 &  640 & 140 & 0.14 & 0.476(14) & 8000  & 0.05 \\
0.37 &  80 &  800 & 1000 &  768 & 275 & 0.22 & 0.45(2)   & 4000  & 0.05 \\
0.38 & 100 & 1000 & 1200 &  640 & 440 & 0.33 & 0.30(4)   & 3400  & 0.05 \\
0.39 &     &      &      &      &     &      &           & 5000  & 0.045 \\
0.40 & 100 & 1000 & 1200 &  960 & 265 & 0.44 & 0.085(24) & 4000  & 0.04\\ 
0.50 &     &      &      &      &     &      &           & 2000  & 0.03 \\
\hline
\end{tabular}
\end{center}
\caption{Parameters for the simulations in the neighbourhood of the onset.
$N_{\text{cfg}}$ is the
total number of configurations used.  $N_{\text{gauge}}$ is an
estimate of the number of independent gauge configurations, based on
the plaquette autocorrelation times.  With respect to fermionic
observables, we assume all configurations to be independent.  $p_-$ is
the fraction of configurations with a negative determinant, while
$\bra r\ket$ is the average reweighting factor, with the sign of the
determinant included. $N_{\text{traj}}$ is the number of HMC trajectories, 
and $\tau$ the average trajectory length.}
\label{tab:sim-params}
\end{table*}

\subsection{Standard fermionic observables}
\label{sec:chiral}

\begin{figure}[tb]
\includegraphics[width=\colw]{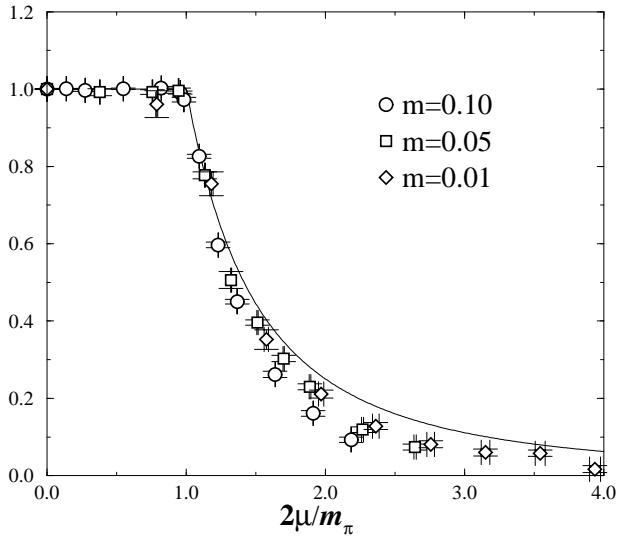}
\caption{Rescaled chiral condensate vs.\ chemical potential
together with the $\chi$PT predictions (\ref{eq:unipbp}).
\label{fig:unipbp}}
\end{figure}

\begin{figure}[htb]
\includegraphics[width=\colw]{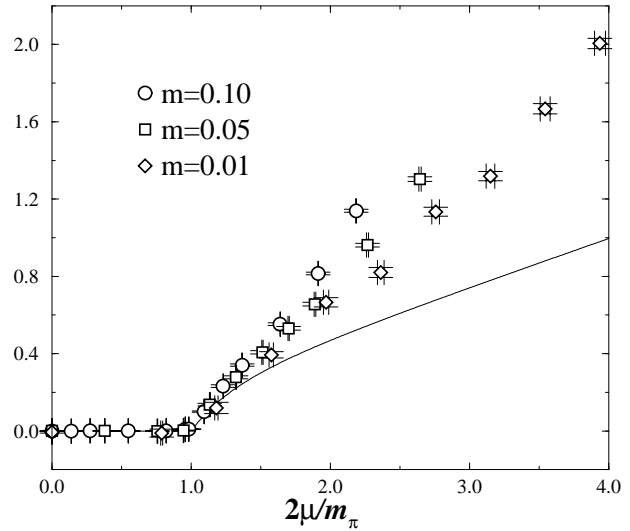}
\caption{Rescaled baryon density vs.\ chemical potential together with
the $\chi$PT predictions (\ref{eq:uniden}).
\label{fig:uniden}}
\end{figure}

\begin{figure}[htb]
\includegraphics[width=\colw]{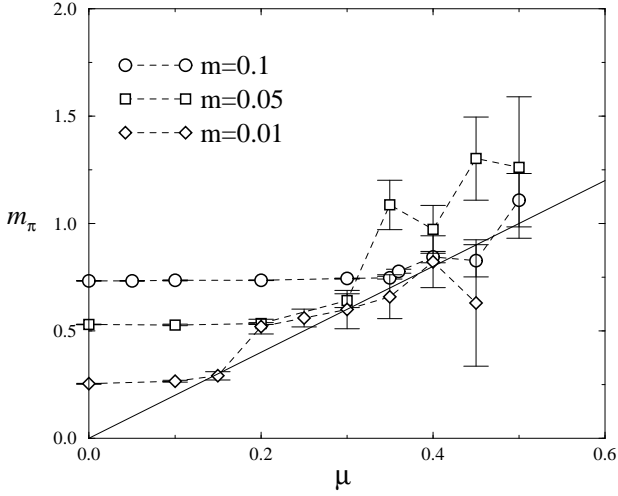}
\caption{$m_\pi$ vs. $\mu$, for the three different quark masses. Also
shown is the line $m_\pi=2\mu$.
\label{fig:pion}}
\end{figure}

First we update our results for the `standard' fermionic
obervables: chiral condensate $\pbp$, fermion density $n$ and pion
mass $m_\pi$.  In \cadj,
we found that the HMC algorithm, which samples only positive
determinant configurations, yields results that agree well with the
$\chi$PT predictions \ckstvz\ for theories with diquark Goldstone modes:
\begin{align}
y =\frac{\pbp}{\pbp_0} &=\begin{cases}
    1&\!x\!<\!1\\\frac{1}{x^2}&\!x\!>\!1
\end{cases}\, ; \label{eq:unipbp} \\
\tilde n =\frac{nm_{\pi0}}{8m\pbp_0} &=\begin{cases}
    0&\!x\!<\!1\\
    \frac{x}{4}\left(1-\frac{1}{x^4}\right)&\!x\!>\!1
     \end{cases}\, ; \label{eq:uniden} \\
m_\pi &=\begin{cases}
    m_{\pi0}&\!x\!<\!1\\
    2\mu&\!x\!>\!1
	\end{cases}\, ;
\label{eq:pion}
\end{align}
where $m$ is the bare quark mass, $x\equiv2\mu/m_{\pi0}$ 
and the subscript 0 denotes the values at
$\mu=0$.  Here we have extended our simulation to higher values of
$\mu$.  The results are shown in figs.~\ref{fig:unipbp},
\ref{fig:uniden} and \ref{fig:pion} for $\pbp$, $n$ and $m_\pi$
respectively.  We see that good qualitative agreement 
with lowest-order $\chi$PT,
in particular for $\pbp$, continues up to $x\simeq2$.
It is also 
worth noting at this stage that the prediction for the onset $x_o=1$, clearly
supported by our data, is stable
to next-to-leading order in $\chi$PT \cite{Splittorff:2001fy}. For
$x\gtrsim2$, however, there is a dispersion between the data for
different values of $m$. This could be explained by 
the appearance of new baryonic
states in the spectrum not described by $\chi$PT
(eg. a spin-1 $qq$ state, or perhaps
even a $qg$ fermion), which should start to populate the ground
state once $\mu\approx m_b/n_q$. Since we expect such states to have a
mass $m_b$ of
order the constituent quark mass, governed by the magnitude of $\pbp_0$, then
for different $m$ the new thresholds should manifest themselves
at different values of the rescaled variable $x$. 
For $x\gtrsim3$ there is also an indication of nonanalytic
behaviour, which may conceivably be a sign of a further phase
transition.  It is also interesting to note that there are no signs of
any saturation effects (i.e., when $n$ approaches its maximum allowed value of 3),
even at our largest $\mu=0.8$.
It is clear, however, that any 
such conclusions remain provisional until data from larger volumes is
available.

\begin{figure}
\includegraphics[height=\colw,angle=-90]{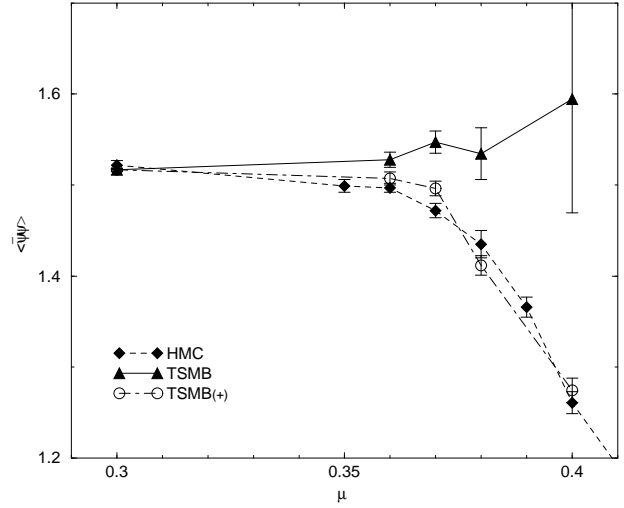}
\caption{The chiral condensate from TSMB and HMC simulations.}
\label{fig:psibarpsi}
\end{figure}

\begin{figure}
\includegraphics[height=\colw,angle=-90]{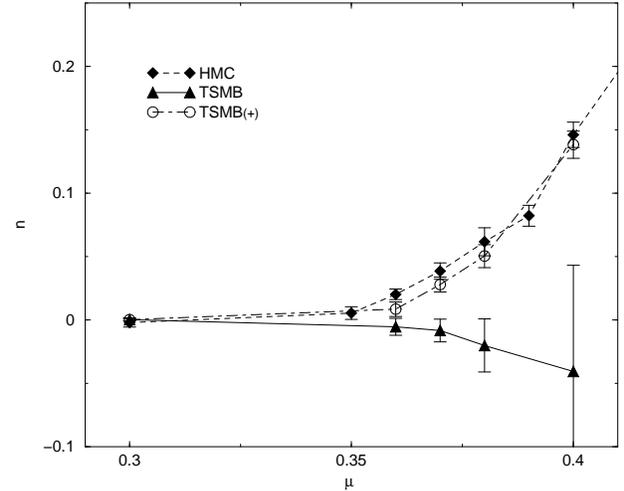}
\caption{The fermion density from TSMB and HMC simulations.}
\label{fig:density}
\end{figure}

In \cadj\ we found indications that the onset transition at
$x\approx1$ may be delayed when the negative configurations are
included using the TSMB algorithm.  Here we have collected additional
data with both algorithms from the region just above this transition
point for $m=0.1$ (see table~\ref{tab:sim-params}).
The results are shown in figs.~\ref{fig:psibarpsi}
and \ref{fig:density} for the chiral condensate and fermion density
respectively.  We can clearly see that the numbers for the positive
determinant sector of TSMB agree well with the HMC data, giving the
same early onset transition, but including the negative determinant
sector brings the total average back so that it is consistent with the
vacuum value --- and inconsistent with the value in the positive
determinant sector.  We conclude from this that, as already indicated,
the positive determinant sector and the full theory represent two
different `physical' models: the former describes a two-flavor theory
with an onset transition at
$\mu_o=m_\pi/2$, while the latter, the true $N=1$ theory, remains in
the vacuum phase for $\mu>m_\pi/2$, and presumably has a 
transition to a normal or superconducting phase at higher $\mu$.
At $\mu=0.4$ we see that the uncertainties in the averages of the full theory
become
large.  This is an effect of the sign problem, which becomes serious
at this point.  Simulating at even larger chemical potential, in the
hope of finding the real onset transition in this model, would be very
demanding and beyond our current resources.

To gain further insight into the nature of the dense phase it is instructive to
consider the quark energy density, given by
\begin{equation}
\epsilon_{\text{phys}}=\langle\bar\chi D_0\chi\rangle
-{1\over4}(3-m\pbp_0)\, .
\end{equation}
The results are shown in table~\ref{tab:enden}, along with the expected energy
density $nm_{\pi0}/2$ of a system with quark number density $n$ consisting
of non-interacting diquark baryons of mass $m_{\pi0}$. The two sets of
numbers are comparable, although to the accuracy we have obtained it is
impossible to assess even the sign of the binding energy in such a picture.
It is also possible to express things in physical units;
since $m_\pi\propto\surd m$ \cadj, and observing that our value
of $m/m_{\pi0}\simeq0.136$ is about 5 times the physical ratio (assuming
$m_u=4$ MeV, $m_\pi=140$ MeV), we conclude that our $m=0.1$
simulation describes a world with $m\simeq90$ MeV,
$m_\pi\simeq670$ MeV. This corresponds to a lattice spacing
$a\simeq0.22$ fm.
At $\mu=0.37$ this gives a quark number density
$n\simeq3.6\text{ fm\,,}^{\!\!-3}$  with a corresponding energy density
$\epsilon\simeq1270\text{ MeV\,fm\,,}^{\!\!\!-3}$ considerably in excess of the
nuclear matter values $n\simeq0.48\text{ fm\,,}^{\!\!\!-3}$ 
$\epsilon\simeq150\text{ MeV\,fm\,.}^{\!\!\!-3}$ Perhaps a more reasonable
comparison, however, is the dimensionless ratio
$\epsilon/(mn)$, which is $\simeq3.8$ at $\mu=0.37$ but has a value closer
to 80 in nuclear matter. This emphasises the large value of the quark mass in
our simulations; a more ``realistic'' simulation would require a separate
calibration, e.g.\ via the vector meson mass.


\begin{table}
\begin{tabular}{rll} \hline
$\mu$ & $\epsilon_{\text{phys}}$ & ${n\over2}m_{\pi0}$\\ \hline
0.20 & 0.0001(20) &  0.0005(10) \\
0.30 & $-0.0047(28)$ &  $-0.0007(12)$ \\
0.35 & $-0.0056(40)$ & \m0.0020(18) \\
0.36 & \m0.0042(31) &  \m0.0074(15) \\
0.37 & \m0.0148(51) &  \m0.0142(24) \\
0.38 & \m0.0188(80) &  \m0.0226(40) \\
0.39 & \m0.0316(70) &  \m0.0302(30) \\
0.40 & \m0.0609(83) &  \m0.0536(37) \\
0.50 & \m0.2688(220) & \m0.1921(88) \\
\hline
\end{tabular}
\caption{Energy density as a function of $\mu$, compared with that of a 
system of non-interacting pions, for simulations with $m=0.1$.}
\label{tab:enden}
\end{table}

In table~\ref{tab:susc} we report the 
susceptibility defined by the integrated pion correlator
\begin{equation}
\chi_\pi=\sum_x\langle\bar\chi\varepsilon\chi(0)\bar\chi\varepsilon\chi(x)
\rangle=\sum_x G_{0,x}(\mu)G^{tr}_{0,x}(-\mu)\, ,
\end{equation}
where $\varepsilon(x)=(-1)^{x_0+x_1+x_2+x_4}$ and $G$ is the fermion propagator,
which is real for adjoint quarks. 
\begin{table}
\begin{tabular}{rllll} \hline
$\mu$ & $\bra\chi_\pi\ket_{\text{TSMB}}$ & $\bra\chi_\pi\ket_+$
& $\bra\chi_\pi\ket_-$ & $\langle\chi_\pi\rangle_{\text{HMC}}$ \\ \hline
0.00 & 45.87(7) \\
0.20 &          & & & 45.80(16)\\
0.30 & 45.76(5) & & & 45.87(22)\\
0.35 &          & & & 45.36(25)\\
0.36 & 45.79(24) & 45.28(21) & 35.32(155) & 45.04(17) \\
0.37 & 46.49(35) & 44.80(22) & 35.02(78)  & 44.18(21) \\
0.38 & 46.23(81) & 42.49(30) & 36.34(56)  & 42.91(45) \\
0.39 &          & & & 40.63(32)\\
0.40 & 46.24(321) & 38.06(34) & 35.56(38) & 37.42(40) \\
0.50 &          & & & 21.15(82)\\
\hline
\end{tabular}
\caption{Pion susceptibility
$\chi_\pi$ from TSMB and HMC simulations.}
\label{tab:susc}
\end{table}
In fact, the tabulated results
come from `non-singlet'
diagrams with connected quark lines only;
disconnected contributions were consistent with zero within large statistical
uncertainties (we also attempted to measure the scalar susceptibility but did
not find a significant signal with our statistics).
The results are consistent, to within an irrelevant normalisation factor of 3,
with the axial Ward identity $m\chi_\pi=\pbp$. Once again the HMC data and the 
positive determinant sector of TSMB are in close agreement, showing a systematic
decrease
of $\chi_\pi$ in the dense phase. Reweighting to the correct ensemble,
however, again
removes the evidence for the onset transition -- this is significant since it
demonstrates that reweighting also gives the expected behaviour for two-point
correlation functions.

\begin{table}
\begin{tabular}{rllll}\hline
$\mu$ & $m_\pi^{\text{TSMB}}$ & $m_\pi^+$
& $m_\pi^-$ & $m_\pi^{\text{HMC}}$ \\ \hline
0.00 & 0.7322(6) & & & \\
0.20 &           & & & 0.7324(7) \\
0.30 & 0.7360(8) & & & 0.7345(16) \\
0.35 &           & & & 0.7504(40) \\
0.36 & 2.00(85)  & 1.35(40)   &   ---    & 0.7580(56) \\
0.37 & 0.686(81) & 0.751(56)  & 2.06(30) & 0.769(10) \\
0.38 & 1.29(65)  & 0.915(128) &   ---    & 0.758(22) \\
0.39 &           & & & 0.819(22) \\
0.40 & 1.19(2.52)&    ---     & 1.16(30) & 0.829(28) \\ \hline
\end{tabular}
\caption{The pion mass $m_\pi$ from TSMB and HMC simulations.}
\label{tab:pion}
\end{table}

The pion mass $m_\pi(\mu)$ is extracted from the temporal decay of the
correlator over timeslices 1--7.  The results are shown in
table~\ref{tab:pion}.  HMC results show that $m_\pi$ is constant 
and equal to $m_{\pi0}$ for $\mu<\mu_o$, and increases for $\mu>\mu_o$ 
in approximate agreement with the $\chi$PT prediction (\ref{eq:pion}); the
fact that the results lie slightly above the $\chi$PT value is probably
attributable to the small volume. It is interesting to contrast this 
behaviour with that observed in two-color QCD with fundamental staggered 
quarks \cite{Kogut:2001na}, in which $m_\pi$ is observed to {\em decrease\/}
once $\mu>\mu_o$. This behaviour is 
predicted by $\chi$PT \ckstvz\ 
for a pseudo-Goldstone meson with a flavor content 
symmetric under the residual global symmetry in the superfluid state, which for
$N=2$ staggered adjoint flavors is Sp(2);
$m_\pi$ should decrease with $\mu$ in a theory with Dyson index $\beta=4$ such
as two-color QCD with staggered fundamental quarks, 
and increase if $\beta=1$ as is the case for the current model.
It becomes very difficult to determine $m_\pi$ from the TSMB simulation
once $\mu>0.3$, so we are not able to draw any significant conclusions
from these data.  However, the general tendency is compatible with what
we observe for $\pbp$ and $n$; namely,
that $m_\pi$ increases (with values compatible with those from
HMC) when only positive determinant configurations are included, while
it remains compatible with the $\mu=0$ value when all configurations
are included.

\subsection{Gauge field quantities}
\label{sec:gauge}

Our results for the plaquette and Wilson loops are given in
table~\ref{tab:gauge-results}.  Beyond the observation that both algorithms 
give consistent results for the plaquette we see no discernible effect over this
range of chemical potentials --- all quantities remain consistent with
their values at $\mu=0$. At larger $\mu$ there is evidence that 
the plaquette starts to fall towards its quenched value 
due to the effects of Pauli blocking 
\cite{Hands:2000ei,Hands:2000yh}.
The deviation of the $\mu=0.37$ values from the rest is probably a
sign of insufficient statistics and/or a too short equilibration time.

\begin{table}
\begin{tabular}{llllll}\hline
$\mathcal{O}/\mu$ & $\bra\mathcal{O}\ket$ & $\bra\mathcal{O}\ket_+$
 & $\bra\mathcal{O}\ket_-$  & $\bra\mathcal{O}\ket_{\text{HMC}}$\\ \hline
$\Box$\\
  0.0 & 0.5667(9) \\
  0.20 &            &            &           & 0.5667(16) \\
  0.30 & 0.5689(12) &            &           & 0.5677(20) \\
  0.35 &            &            &           & 0.5689(20) \\
  0.36 & 0.5687(20) & 0.5684(19) & 0.5623(27)& 0.5644(30) \\
  0.37 & 0.5584(15) & 0.5583(14) & 0.5572(21)& 0.5642(35) \\
  0.38 & 0.5696(15) & 0.5688(12) & 0.5676(13)& 0.5677(35) \\
  0.39 &            &            &           & 0.5685(30) \\
  0.40 & 0.5579(38) & 0.5597(18) & 0.5602(19)& 0.5676(40) \\ 
  0.50 &            &            &           & 0.5634(50) \\
\hline
$\Box_s$ \\
 0.0 & 0.5667(9) \\
 0.30 & 0.5688(13)\\
 0.36 & 0.5686(19) & 0.5683(19) & 0.5625(30) \\
 0.37 & 0.5595(15) & 0.5590(15) & 0.5566(22) \\
 0.38 & 0.5687(18) & 0.5682(14) & 0.5667(15) \\
 0.40 & 0.5583(43) & 0.5600(20) & 0.5602(21) \\ \hline
$W_{11}$ \\
 0.0 & 0.5668(9) \\
 0.30 & 0.5687(12) \\
 0.36 & 0.5691(22) & 0.5688(22) & 0.5617(26) \\
 0.37 & 0.5574(14) & 0.5575(14) & 0.5579(21) \\
 0.38 & 0.5696(15) & 0.5688(13) & 0.5676(16) \\
 0.40 & 0.5575(41) & 0.5593(17) & 0.5599(18) \\ \hline
$W_{12}$ \\
 0.0 & 0.3426(14) \\
 0.30 & 0.3452(16) \\
 0.36 & 0.3463(33) & 0.3459(32) & 0.3375(37) \\
 0.37 & 0.3291(19) & 0.3292(19) & 0.3294(29) \\
 0.38 & 0.3474(22) & 0.3462(20) & 0.3443(21) \\
 0.40 & 0.3299(59) & 0.3320(25) & 0.3326(28) \\ \hline
$W_{21}$ \\
 0.0 & 0.3422(14) \\
 0.30 & 0.3451(18) \\
 0.36 & 0.3458(35) & 0.3543(34) & 0.3356(36) \\
 0.37 & 0.3289(22) & 0.3292(21) & 0.3308(28) \\
 0.38 & 0.3451(26) & 0.3453(21) & 0.3457(23) \\
 0.40 & 0.3296(60) & 0.3321(25) & 0.3328(28) \\ \hline
$W_{22}$ \\
 0.0 & 0.1408(15) \\
 0.30 & 0.1439(18) \\
 0.36 & 0.1460(35) & 0.1456(35) & 0.1377(39) \\
 0.37 & 0.1268(22) & 0.1269(21) & 0.1279(28) \\
 0.38 & 0.1478(30) & 0.1477(24) & 0.1474(25) \\
 0.40 & 0.1303(68) & 0.1318(31) & 0.1323(33) \\ \hline
 $L$ \\ 
 0.00 & \!\!-0.003(3) \\
 0.30 & \!\!-0.005(4) \\
 0.36 &  0.015(9)  &  0.014(9) & \!\!-0.002(10) \\
 0.37 & \!\!-0.002(4)  &  0.000(4) &  0.011(7)  \\
 0.38 &  0.010(9)  &  0.008(7) &  0.004(10) \\
 0.40 & \!\!-0.005(18) & \!\!-0.010(6) & \!\!-0.11(6)  \\ \hline
\end{tabular}
\caption{Results for the gauge observables: the plaquette $\Box$,
spatial plaquette $\Box_s$, timelike Wilson loops $W_{ij}$, and Polyakov loop
$L$.}
\label{tab:gauge-results}
\end{table}

The Polyakov loop $L$ on small lattices 
in the dense phase of two-color QCD with fundamental quarks
was found to be small but non-zero in \cite{Kogut:2001na}.
Our results for the average Polyakov loop are also given in
table~\ref{tab:gauge-results}.  We can see that it remains zero everywhere,
and conclude that there is no sign of any deconfinement transition
either in the positive determinant sector or in the full theory.

\subsection{Diquark condensation}
\label{sec:diquark}

A natural explanation of the agreement between the HMC and positive
determinant results and $\chi$PT predictions is
that the positive determinant sector mimics a theory with two flavors:
one quark and one `conjugate quark' \cite{Stephanov:1996ki}.  We may
cast further light on this by studying the behaviour of the diquark
condensate (\ref{eq:qq3}) expected in the two-flavor theory.  This
condensate may be determined by introducing a diquark source term in
the action, which now describes two flavors,
\begin{equation}
S[j] = S + j\sum_x qq_{\bf3}(x) \, ,
\label{eq:diquark-source}
\end{equation}
extracting the condensate 
$\langle qq_{\bf3}(j)\rangle\!=\!V^{-1}\partial\ln Z[j]/\partial j$,
and extrapolating the results to $j=0$ \cite{Morrison:1998ud}.
Here, we have only performed
`partially quenched' measurements --- i.e., we have included a nonzero diquark
source $j$ in the measurement of observables from configurations
generated at $j=0$.

\begin{figure}
\includegraphics[width=\colw]{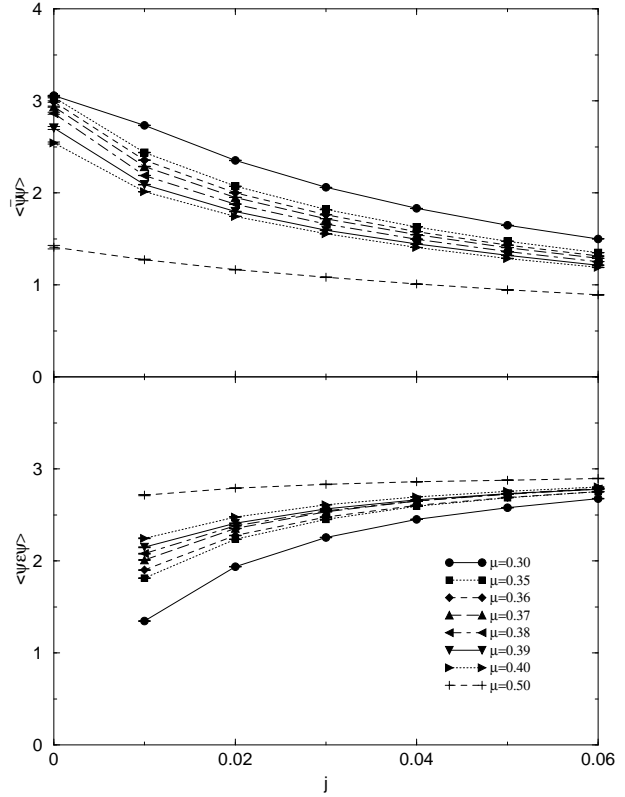}
\caption{The chiral and diquark condensates as a function of the
chemical potential $\mu$ and the diquark source $j$, from HMC.}
\label{fig:cond-src}
\end{figure}

In figure~\ref{fig:cond-src} we show the chiral and two-flavor diquark
condensates from our HMC simulations as a function of $j$ for a number
of values of the chemical potential.  Note that on a finite system
at $j=0$ the diquark
condensate is identically zero.  We have therefore
excluded this point from our plot.

\begin{figure}
\includegraphics[width=\colw]{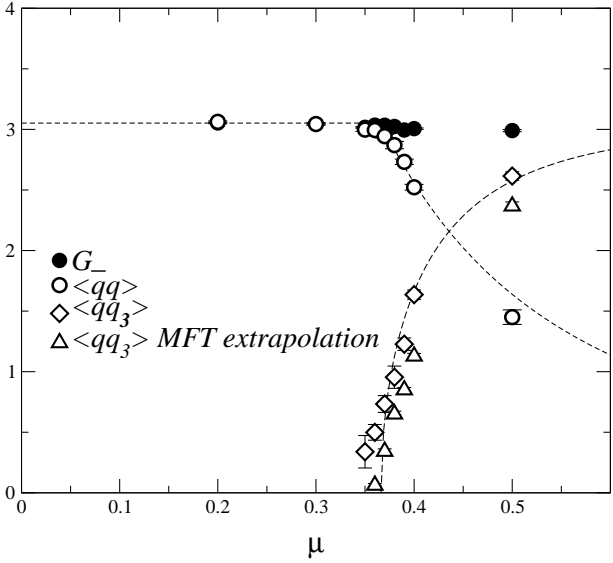}
\caption{$\pbp$, $\langle qq_{\bf3}\rangle$ 
and $G$ 
versus $\mu$, together with predictions from leading order $\chi$PT. 
``MFT extrapolation'' denotes points from fits to
(\ref{eq:diq-extrap})
}
\label{fig:conds}
\end{figure}

The extrapolation $j\to0$ needs some discussion.
To leading order in $\chi$PT the relation between chiral and diquark 
condensates, $m$, and $j$ is \ckstvz
\begin{equation}
x^2\pbp\langle qq_{\bf3}\rangle=\frac{\pbp_0}{\sqrt{1+\frac{j^2}{m^2}}}\left[
\bra qq_{\bf3}\ket-\frac{j}{m}\pbp\right],
\label{eq:uniqq}
\end{equation}
together with the constraint
\begin{equation}
\pbp_0^2=\pbp^2+\langle qq_{\bf3}\rangle^2.
\label{eq:pythag}
\end{equation}
If we assume $j\ll m$ and ignore quadratic terms, then at
at the critical point $x_o=1$ the resulting cubic equation yields
$\langle qq_{\bf3}\rangle/\pbp_0
=^3\!\!\!\!\!\surd2({j\over m})^{1\over3}-{2\over3}
{j\over m}$.
We have therefore tried to fit the data using the form
\begin{equation}
\bra qq_{\bf3}(j)\ket = A + Bj^{1\over3}+Cj\, .
\label{eq:diq-extrap}
\end{equation}
The results of the extrapolation are shown in fig.~\ref{fig:conds}. Positive 
values for the coefficient $A$ were found for $\mu\geq0.36$, and the quality of
the fit improved as $\mu$ increased. At $\mu=0.36$ the values for the 
coefficients
were $A=0.062(15)$, $B=9.37(7)$, $C=-16.1(2)$, the latter to be compared with
$\chi$PT predictions at $x_o=1$ of 8.28 and -20.4 respectively.
Although the quality of the fits 
improves as $\mu$ increases, the curvature due to the $j^{1\over3}$ term
is observed 
to manifest itself principally at values of $j<0.01$, i.e., below where we 
have data.
We therefore suspect that fits to (\ref{eq:diq-extrap}) underestimate 
$\langle qq_{\bf3}(j=0)\rangle$ for $\mu>\mu_o$.  This is in line with
the solutions of (\ref{eq:uniqq},\ref{eq:pythag}) at $x>1$:
extrapolating the solutions of (\ref{eq:uniqq}) at $j>0$ to $j=0$
using (\ref{eq:diq-extrap}) always yields a lower value than the
actual solution at $j=0$,
\begin{equation}
\bra qq_{\bf3}\ket = \pbp_0\sqrt{1-\frac{1}{x^4}}\, .
\label{eq:unidiq}
\end{equation}

We can also see from fig.~\ref{fig:cond-src} that the behaviours of
$\pbp$ and $\bra qq_{\bf3}\ket$ are strongly anti-correlated.  This
is a result of the two quantities being connected by a $\U(4)_f$
transformation \cadj,
\begin{gather}
X=\begin{pmatrix}\chi_o\\ \bar\chi_o^{tr}\end{pmatrix}\to VX \quad
\bar{X}=\begin{pmatrix}\bar\chi_e&\chi_e^{tr}\end{pmatrix}\to
\bar{X}V^{\dagger} \, , \\
\intertext{with}
V=\half\begin{pmatrix}\phantom{i}P&iP^{tr}\\
iP&\phantom{i}P^{tr}\\\end{pmatrix} \quad\text{where}\quad
P=\begin{pmatrix}1&-1\\1&\phantom{-}1\\\end{pmatrix}.
\end{gather}
The superfluid transition in this model is expected to be realised by
the chiral condensate `rotating' into the diquark condensate, in such
a way that the constraint (\ref{eq:pythag}) is maintained.
This has been shown to be the case to leading order in $\chi$PT
\ckstvz.  At next to leading order, however, some dependence on $\mu$ and $j$
is expected \cite{Splittorff:2001fy}. 

We have therefore tried a second method in which the quantity 
$G^2=\pbp^2+\langle qq_{\bf3}\rangle^2$ measured for $j>0$
is extrapolated to $j=0$
using a second order polynomial. Fig.~\ref{fig:G-extrap} 
shows that $G^(j,\mu)$ exhibits relatively little variation with either $j$ or
$\mu$, implying that the NLO $\chi$PT corrections are small. Note that 
since $\langle qq_{\bf3}(j=0)\rangle\equiv0$, 
the extrapolated $G^(0)$ lies
considerably above the numerical data for $\mu>\mu_o$. We now use $G^(0)$, 
which appears to vary very little across the transition at $\mu=\mu_o$,
together with $\pbp$ to extract $\langle qq_{\bf3}\rangle$; the results are
plotted in fig.~\ref{fig:conds}. 
In this case difficulties associated 
with taking the difference of two fluctuating quantities means that the method
is probably not accurate for $\langle qq_{\bf3}\rangle$
in the immediate vicinity of the critical point; for
$\mu\geq0.37$, however, the agreement with the $\chi$PT predictions
({\ref{eq:unipbp},\ref{eq:unidiq}) is striking.

\begin{figure}
\bigskip\bigskip
\includegraphics[width=\colw]{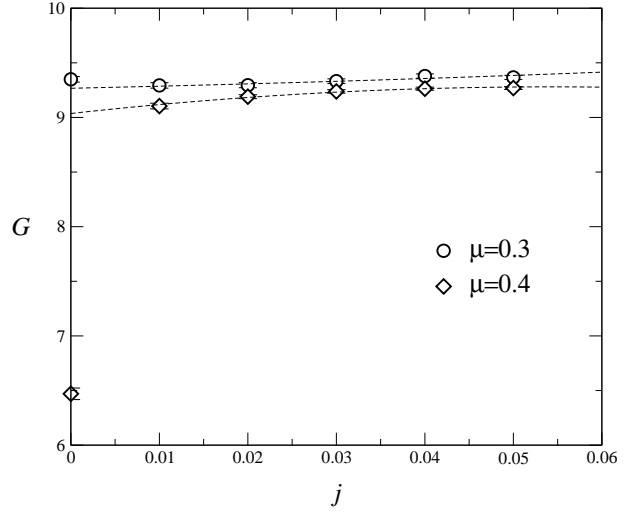}
\caption{The quantity $G^2(j)$ for values of $\mu$ in either phase. Lines show
the quadratic extrapolation to $j=0$.
}
\label{fig:G-extrap}
\end{figure}

In figure~\ref{fig:diquark-tsmb} we compare the HMC results for 
$\langle qq_{\bf3}\rangle$ at $\mu=0.38$ with TSMB numbers.  Since the
superfluid transition is not present for $N=1$, the diquark condensate
should be suppressed, and indeed the 
data from the full theory lie lower and exhibit a stronger negative
curvature than the positive determinant data.  It is, however, not
possible to conclude from these data whether the one set will
extrapolate to zero and the other to a non-zero value, especially
taking into account the problems discussed above.
\begin{figure}
\includegraphics[height=\colw,angle=-90]{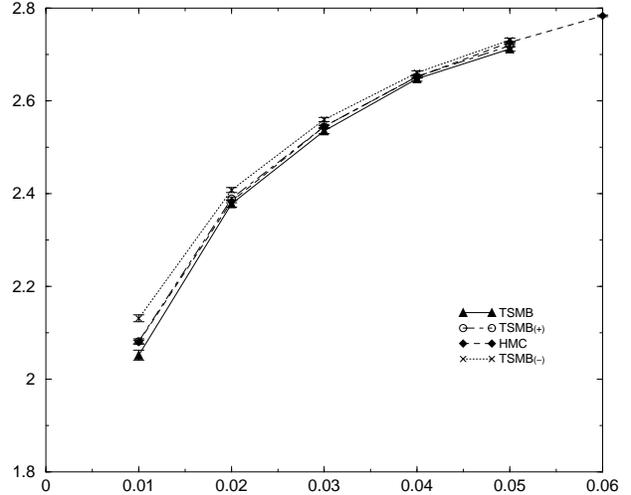}
\caption{The diquark condensate as a function of the diquark source
$j$ at $\mu=0.38$, from TSMB and HMC.}
\label{fig:diquark-tsmb}
\end{figure}

\begin{figure*}
\begin{center}
\includegraphics[width=0.75\textwidth]{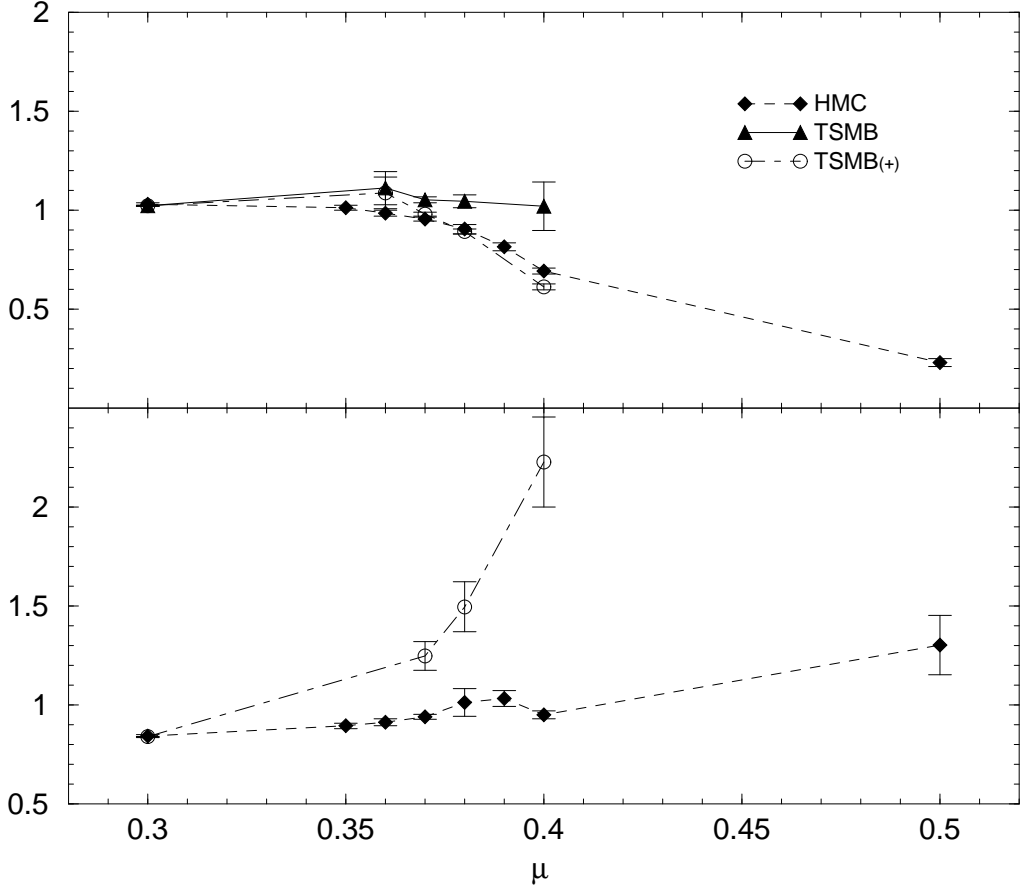}
\end{center}
\caption{The superconducting (top) and superfluid (bottom) diquark
susceptibilites as a function of the chemical potential $\mu$, from
HMC and TSMB.}
\label{fig:diq-susc}
\end{figure*}

We have also computed the following four-point functions, which we refer to as 
superfluid and superconducting ``susceptibilities'':
\begin{align}
\chi_{\bf3} & =
 \bra\bar\chi(x)\varepsilon\bar\chi^{tr}(x)
 \chi^{tr}(x)\varepsilon\chi(x)\ket\, ,
 \label{eq:susc3} \\
\chi_{sc} & = \frac{1}{3}\bra\bar\chi(x) t^i\bar\chi^{tr}(x)
 \chi^{tr}(x)t^i\chi(x)\ket \, . \label{eq:suscsc}
\end{align}
Both may be computed without introducing a diquark source term in the
action \cite{Hands:1998kk,Morrison:1998ud}, 
and are related to the superfluid and superconducting diquark
condensates of eqs.~(\ref{eq:qq3}) and (\ref{eq:qqsc}) respectively.
Both are scalar objects corresponding to the local component of a
diquark susceptibility $\chi_{qq}=\sum_x\langle qq(0)\bar q\bar q(x)\rangle$
which we might expect to increase significantly if condensation occurs.
Note, however, that $\chi_{sc}$ in (\ref{eq:suscsc}) is the only such
contribution which is gauge invariant and hence a possible signal of color
superconductivity in a simulation (to be compared with $\Phi^\dagger\Phi$ in a
Higgs model). To be able to compare signals for condensation between channels 
we have therefore chosen the
local operators (\ref{eq:susc3},\ref{eq:suscsc}) over the full susceptibilities.

Figure~\ref{fig:diq-susc} shows the superfluid and superconducting
diquark susceptibilities as a function of chemical potential from our
HMC and TSMB simulations.  In the low density regime $\chi_{\bf3}$ and
$\chi_{sc}$ show little variation with $\mu$ and are roughly equal.  
Once we enter the superfluid phase for $\mu>\mu_o$, however, 
signals for both observables become increasingly dominated by very sharp peaks
of amplitude $O(10 - 20)$, the distinction being that for $\chi_{\bf3}$ the 
peaks are all positive whereas for $\chi_{sc}$ they occur with either sign.
Such peaks are characteristic of localised small-eigenvalue modes of 
$D{\!\!\!\!/\,}$, of which there are many in the dense phase 
\cite{Hands:2000ei}.
For $\mu>\mu_o$, in the positive determinant sector $\chi_{sc}$ 
decreases. We
interpret this as an effect of Pauli blocking or phase space
suppression: as the ground state is filled up by fermions, there is
less phase space left to accommodate 
the fermion loops that contribute to $\chi_{sc}$.

When the sign is taken into account,
this effect, like other effects of the superfluid transition,
disappears.  Perhaps surprisingly, this is in fact the cleanest signal
we have of the effect of the sign of the determinant on observables.
By way of contrast, $\chi_{\bf3}$ shows a small increase as the onset
transition is reached, as one would expect as a result of a Goldstone mode
once a superfluid condensate forms.
However,
there is a large deviation between $\chi_{\bf3}$ as measured
in HMC and TSMB(+); this is the only
observable we have examined in which the two algorithms significantly
fail to agree. One possible explanation is that Goldstone physics is 
extremely sensitive to small eigenmodes of $D{\!\!\!\!/\,}$. In HMC such modes 
are responsible for large force terms and reduced acceptance rates; 
moreover the sampling may not be correct due to the breakdown of ergodicity
\cadj. Hybrid algorithms have produced disparate results for susceptibilities
in the dense phase \cite{Morrison:1998ud,Hands:1999md}.
TSMB, in contrast, tends to 
over-sample configurations with small modes, 
correcting this by reweighting factors with modulus less than 1.  In
fact, in this case some of the eigenvalues and reweighting factors are
very small, and therefore problems with numerical accuracy cannot be
excluded.  It is 
perhaps not surprising that the algorithms disagree in this exacting regime.
We were not able to obtain any result for
the overall superfluid susceptibility in TSMB (i.e., with the negative
determinant sector included) due to huge fluctuations in the signal.

\section{Conclusions}
\label{sec:conclusions}

We have studied two-color QCD with one flavor of staggered quark in
the adjoint representation. We have employed two different simulation
algorithms, and have continued to gain insight into the optimal tuning of the
TSMB algorithm in the high density regime. This is the preferred algorithm
not only because as highlighted in \cadj\ it is capable of maintaining
ergodicity via its ability to change the determinant sign once $\mu>\mu_o$, 
but also because it more effectively samples small eigenmodes, which are
important in the presence of a physical Goldstone excitation. 
We find that the positive
determinant sector behaves like a two-flavor model, and exhibits good
agreement with chiral perturbation theory predictions for such a
model in the regime $1<x<2$.  
At higher chemical potentials there are preliminary indications of a breakdown
of $\chi$PT, and possible signs of a further phase
transition. However, data from larger volumes and smaller bare quark masses 
would be needed to make these observations definitive. 

Above the onset transition in the positive determinant sector, we have
successfully obtained a 
signal for a non-zero two-flavor diquark condensate $\langle
qq_{\bf3}\rangle$, indicating a
superfluid ground state for $\mu>m_\pi/2$. The chiral condensate
rotates into this diquark condensate, in good quantitative agreement with 
the behaviour predicted by $\chi$PT.  This feature also enabled us to
achieve reasonable control
over the necessary $j\to0$ extrapolation.  We also find that the
superfluid susceptibility (\ref{eq:susc3}) increases, while the
superconducting susceptibility (\ref{eq:suscsc}) decreases for $\mu>\mu_o$.
For the former there are indications of a lack of consistency between HMC and
TSMB algorithms, indicative of the two methods' differing treatment of small
eigenmodes.
The
latter may be interpreted as an effect of phase space suppression.
Unfortunately we have seen no evidence for a superconducting condensate 
$\langle qq^i_{sc}\rangle\not=0$, 
one of our original motivations for studying the model.

When the negative determinant configurations are included in the
measurement, the onset transition and diquark condensation disappear.
This is what we would expect for the one-flavor model and is consistent with the
Exclusion Principle. 
There is much stronger
evidence for this scenario
than that presented in our previous paper \cadj, providing
a conclusive demonstration, should any still be needed, that the determinant 
sign plays a decisive role in determining the ground state of systems with
$\mu\not=0$.
Unfortunately,
the severity of the sign problem means we have not been able to locate
the real onset transition for this model.

\begin{acknowledgement}

This work is supported
by the TMR network ``Finite temperature phase transitions in particle physics''
EU contract ERBFMRX-CT97-0122. 
Numerical work was performed using a Cray T3E
at NIC, J\"ulich and an SGI Origin2000 in Swansea.
We are grateful to Kim Splittorff and Don Sinclair for stimulating discussions.
\end{acknowledgement}


\end{document}